\documentstyle[aps,prl,epsf,floats,twocolumn]{revtex}
\begin{document}
\draft
\twocolumn[\hsize\textwidth\columnwidth\hsize\csname @twocolumnfalse\endcsname
\title{Minimum Metallic Mobility in a Two-Dimensional Electron Gas}
\author{Dragana Popovi\'{c}$^{1}$, A.B.\ Fowler$^{2}$, and S.\ Washburn$^{3}$}
\address{$^{1}$Department of Physics, City College of the City University of 
New York, New York, NY 10031 and \\ National High Magnetic Field Laboratory, 
Florida State University, Tallahassee, FL 32306 \\ 
$^{2}$IBM Research Division, T.J.\ Watson Research Center, Yorktown
Heights, NY 10598 \\
$^{3}$Dept.\ of Physics and 
Astronomy, The University of North Carolina at Chapel Hill, Chapel Hill, NC 
27599 }
\date{\today}
\maketitle

\begin{abstract}

We report the observation of a metal-insulator transition in a 
two-dimensional electron gas in silicon.  By applying substrate bias, we have
varied the mobility of our samples, and observed the creation of the
metallic phase when the mobility was high enough 
($\mu${\scriptsize $\stackrel{\textstyle _>}{_\sim}$}1~m$^2$/Vs), 
consistent with the assertion that this transition is driven by 
electron-electron interactions.  
In a perpendicular magnetic field, the magnetoconductance is positive in the 
vicinity of the transition, but negative elsewhere.  Our experiment suggests
that such behavior results from a decrease of the spin-dependent part of the 
interaction in the vicinity of the transition.

\end{abstract}

%
%
%
]

Following the development of the scaling theory of localization for 
non-interacting electrons~\cite{gang}, it has been widely believed that there
is no metal-insulator transition (MIT) in two dimensions, in agreement
with early experiments~\cite{weakloc} on a two-dimensional electron gas (2DEG)
in low-mobility (0.2--0.65~m$^{2}/$Vs) Si metal-oxide-semiconductor 
field-effect transistors (MOSFETs).  However, recent experiments~\cite{Krav}
on high-mobility (2--7~m$^{2}/$Vs) MOSFETs have provided clear evidence for 
the existence of a MIT in this 2D system, raising speculation that this
transition is driven by electron-electron interactions.  In an entirely
different experiment~\cite{dp} on mesoscopic Si MOSFETs of comparable quality 
(peak mobility $\sim$2~m$^{2}/$Vs), non-monotonic behavior of several
characteristic energy scales was found in the transition region, in clear
contradiction with general considerations based on non-interacting 
models~\cite{gang}.  The apparent conflict between these experiments and the
scaling theory of localization has been resolved recently by formulating a
scaling theory for interacting electrons~\cite{newgang}, which shows that the 
existence of a 2D MIT does not contradict any general scaling idea or
principle.  So 
far, however, it has not been shown experimentally how the results of 
Ref.~\cite{Krav} can be reconciled with those obtained on low-mobility 
samples~\cite{weakloc}.  It is exactly this issue that our current work 
resolves.

We present the results obtained on a 2DEG in Si MOSFETs with
a peak mobility of 0.5--0.8~m$^{2}/$Vs at 4.2~K.  In these samples, all 
electronic states are localized, in agreement with early 
studies~\cite{weakloc}.  By applying the substrate bias, however, we have been
able to increase the peak mobility to $\approx$1~m$^{2}/$Vs, and observe the 
metal-insulator transition.  As shown
below, the resulting scaling behavior of the conductivity with temperature $T$
is in excellent agreement with the results of Ref.~\cite{Krav}.  Therefore,
in this way one can study, on a {\em single} sample, the emergence of the
metallic phase once the mobility becomes sufficiently high at low carrier
densities, where the Coulomb interaction is an order of magnitude greater than
the Fermi energy.  In addition, our magnetoconductance measurements in a weak
perpendicular field show that the spin-dependent part of the electron-electron
interaction decreases sharply at the transition.

Our measurements were carried out on n-channel MOSFETs fabricated on the (100)
surface of Si doped at $\approx 8.3\times 10^{14}$ acceptors/cm$^{3}$ with 
435~\AA\, gate oxide thickness and oxide charge $\approx 3\times 
10^{10}$cm$^{-2}$.  The samples had a Corbino (circular) geometry with the 
channel length $L=0.4$~mm and width $W=8$~mm.  Conductance $G$ was measured as
a function of gate voltage $V_{g}$ (proportional to carrier density $n_{s}$) 
at temperatures $1.2 < T < 4.2$~K, using a lock-in at a frequency of 
$\sim 100$~Hz and an excitation voltage of 0.3~mV.

\begin{figure}[t]
\epsfxsize=3.4in \epsfbox{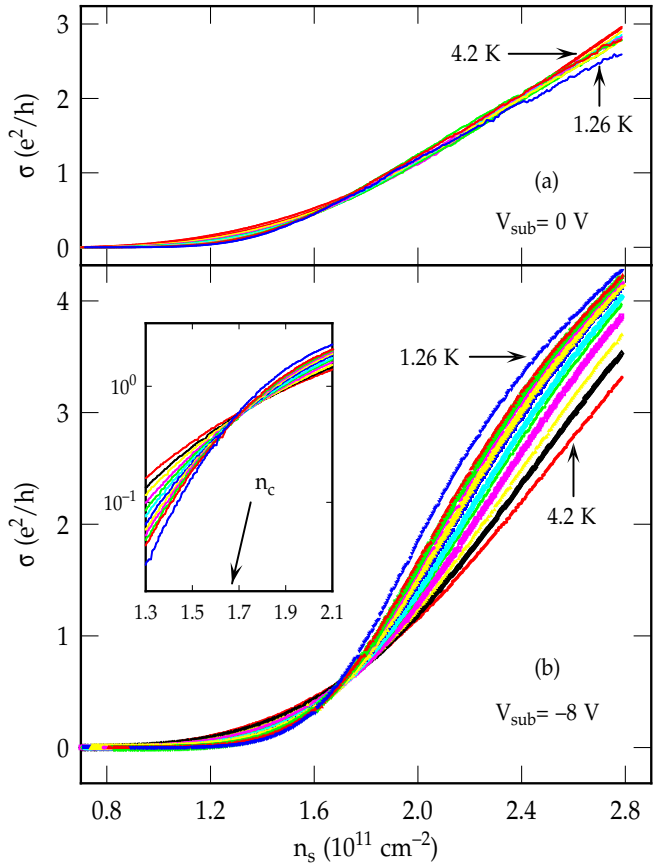}\vspace{5pt}
\caption{Conductivity $\sigma$ of sample \#5 as a function of $n_{s}$ for 
$T=4.2, 3.6, 3.2, 2.8, 2.5, 2.3, 2.08, 1.93, 1.79, 1.67, 1.57, 1.26$~K and (a)
$V_{sub}=0$~V, (b) 
$V_{sub}=-8$~V.  In (a),
$\sigma$ decreases for all $n_{s}$ as $T$ is lowered.  In (b), $\sigma$
increases as $T$ goes down for all $n_{s}>n_{c}$.  The inset shows the same
data around $n_{c}$ with $\sigma$ on a logarithmic scale.
\label{dhva5cond}}
\end{figure}

Fig.~\ref{dhva5cond}(a) shows the conductivity $\sigma =GL/W$
for one of the samples (sample \#5), as a function of $n_{s}$ for 
several temperatures and with no substrate bias $V_{sub}$ applied.  As $T$ is 
lowered, $\sigma$ decreases for both low and high $n_{s}$, indicating 
insulating behavior.  The temperature dependence is weaker at intermediate
values of $n_{s}$.

Application of a substrate bias in Si MOSFETs at a given $n_{s}$ changes the
electric field at the Si-SiO$_{2}$ interface.  As a result, the average
position of the 2D electrons with respect to the interface is changed as well
as the splitting between the subbands~\cite{AFS}.  (At low temperatures, all
electrons populate only the lowest subband and, therefore, are confined to 
motion only in the plane parallel to the Si-SiO$_{2}$ interface.)  The reverse
$V_{sub}$ moves the electrons closer to the interface.
In some samples, especially those with lightly doped substrates such
as ours, where the subband splitting is comparatively small, the mobility 
$\mu=\sigma/n_se$ is enhanced~\cite{Alan}.  This has been attributed to
the reduction of scattering of the 2DEG by electrons that occupy very long 
band 
tails associated with upper subbands and act as additional scattering centers.
The reverse $V_{sub}$ increases the subband splitting and makes this 
scattering mechanism less important.  It may also reduce the average effective
mass since the upper subband electrons are heavier.  In addition, the reverse
$V_{sub}$ reduces the average spatial extent $\Delta z$ of the inversion layer 
charge density in the direction perpendicular to the interface (typically,
$\Delta z\approx 20-30$~\AA~\cite{AFS}).

We find, for example, that $V_{sub}=-1$~V leads to a slight increase of $\mu$ 
and the merging of different temperature curves over a wide range of
$n_{s}$, roughly between 1.8 and $2.4\times 10^{11}$cm$^{-2}$.  At the highest
$n_{s}$, the temperature dependence of $\sigma$ still indicates insulating
behavior.  The reason for the merging at intermediate values of $n_{s}$
becomes clear upon the application of an even higher (reverse) $V_{sub}$,
as shown in Fig.~\ref{dhva5cond}(b) for $V_{sub}=-8$~V: the temperature
dependent behavior of $\sigma$ is now {\em reversed} for all $n_{s}$ above
some critical electron density $n_{c}$ 
($n_{c}=(1.67\pm 0.02)\times 10^{11}$cm$^{-2}$ for sample \#5), as expected
for metallic behavior.  For $n_{s}< n_{c}$, the 2DEG exhibits insulating
behavior as before.  We have obtained similar results on another sample
(sample \#1), where $V_{sub}=-9$~V was used to raise the peak mobility to
1~m$^{2}/$Vs, and study the metal-insulator transition at 
$n_{c}=(1.65\pm 0.02)\times 10^{11}$cm$^{-2}$.

\begin{figure}
\epsfxsize=3.4in \epsfbox{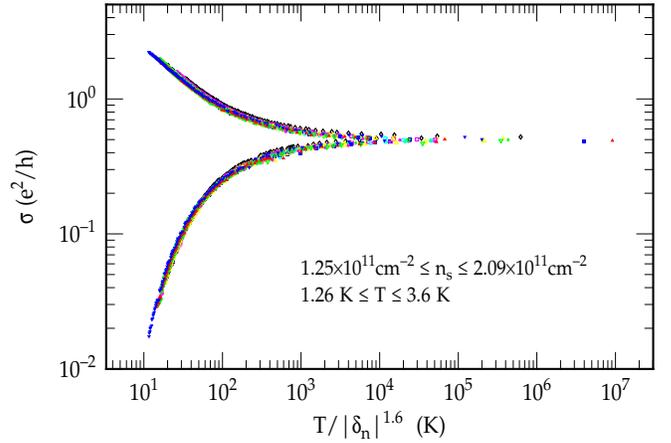}\vspace{5pt}
\caption{Scaling of conductivity with temperature for sample \#5, using the
data shown in Fig.~1(b) in the $n_{s}$ and $T$ ranges given on
the plot.  
\label{bzeroscaling5}}
\end{figure}

The results displayed in Fig.~\ref{dhva5cond}(b) are very similar to those
reported in Ref.~\cite{Krav}, where it was also found that the resistivity
$\rho =1/\sigma$ scales with temperature near the transition according to
\begin{equation}
\label{eq1}
\rho (T,n_{s})=f(|\delta_{n}|/T^{1/z\nu})=\rho (T/T_{0}),
\end{equation}
with a single parameter $T_{0}$ that is the same function of 
$\delta _{n}\equiv (n_{s}-n_{c})/n_{c}$ on both the metallic and the
insulating side of the transition, $T_{0}\propto |\delta _{n}|^{z\nu}$.
Here $z$ is the dynamical exponent, and $\nu$ is the correlation length
exponent.  
The results of this scaling for the data in Fig.~\ref{dhva5cond}(b)
are shown in Fig.~\ref{bzeroscaling5}.  All the data collapse onto two
branches, the upper
one for the metallic side of the transition, and the lower one for the 
insulating side.  The best collapse is found for $z\nu =1.6\pm 0.1$, 
where quite a wide range of electron densities 
(with $|\delta_{n}|$ up to 0.25) was used.  For sample \#1, we find that 
$z\nu=1.5\pm 0.1$.  These results are in a remarkable agreement with 
experiments on high-mobility MOSFETs~\cite{Krav}.  We find that the critical 
conductivity $\sigma_{c}=0.5 e^{2}/h=e^{2}/2h$ for sample \#5, and 
$\sigma_{c}=0.65 e^{2}/h\approx e^{2}/1.5h$ for sample \#1, consistent with 
the reported~\cite{Krav} values and suggesting that the critical conductivity 
has a value around $e^{2}/2h$.

\begin{figure}
\epsfxsize=3.4in \epsfbox{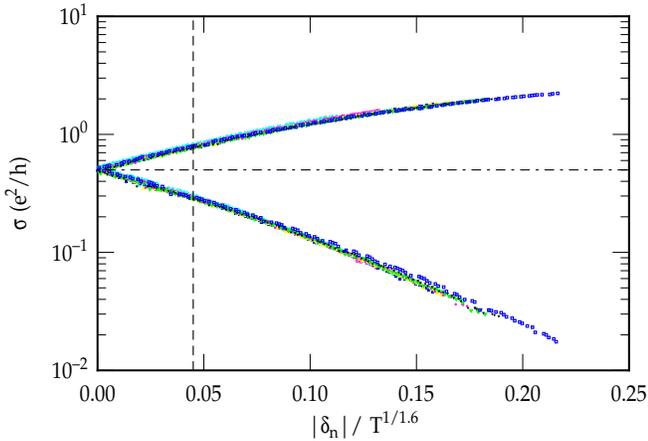}\vspace{5pt}
\caption{Temperature dependence of $\sigma$ for sample \#5, using the
data shown in Fig.~2.  Vertical and horizontal dashed lines are visual aids:
the horizontal line shows the value of $\sigma_c$, and the vertical line shows
the upper limit of the range of $|\delta_n|/T^{1/1.6}$ within which the 
symmetry is observed.
\label{linbzero5}}
\end{figure}

The scaling theory predicts~\cite{newgang} that, to leading order, the 
temperature dependence of the conductivity in the critical region associated 
with the metal-insulator transition will be given by
\begin{equation}
\label{eq2}
\sigma (\delta_n,T)=\sigma_{c}\exp\left(A \delta_{n}/T^{\frac{1}{z\nu 
}}\right),
\end{equation}
where $A$ is an unknown constant.  In other words,
the temperature dependence of $\sigma$ is predicted to be exponential on {\em 
both} insulating and metallic sides of the transition, and with
the same exponent.  This symmetry is expected to 
hold only for $T>T_0(\delta_n)$, i.~e. the quantum critical region is defined 
by the temperature scale $T_0$.  The data presented in 
Fig.~\ref{bzeroscaling5} are shown again in Fig.~\ref{linbzero5} as a function 
of $|\delta_n|/T^{1/z\nu}$, with $z\nu =1.6$.  They are consistent with the
theory, and we observe the symmetry for $|\delta_n|/T^{1/1.6}$ of up to 0.045.
This value defines $T_{0}(|\delta_n|)$, and we find that 
$T_{0}=143|\delta_n|^{1.6}$.  For the range of temperatures of up to 3.6~K, for
which scaling works, this means that the quantum critical region extends at
most up to $|\delta_n|=0.1$.

\begin{figure}[t]
\epsfxsize=3.4in \epsfbox{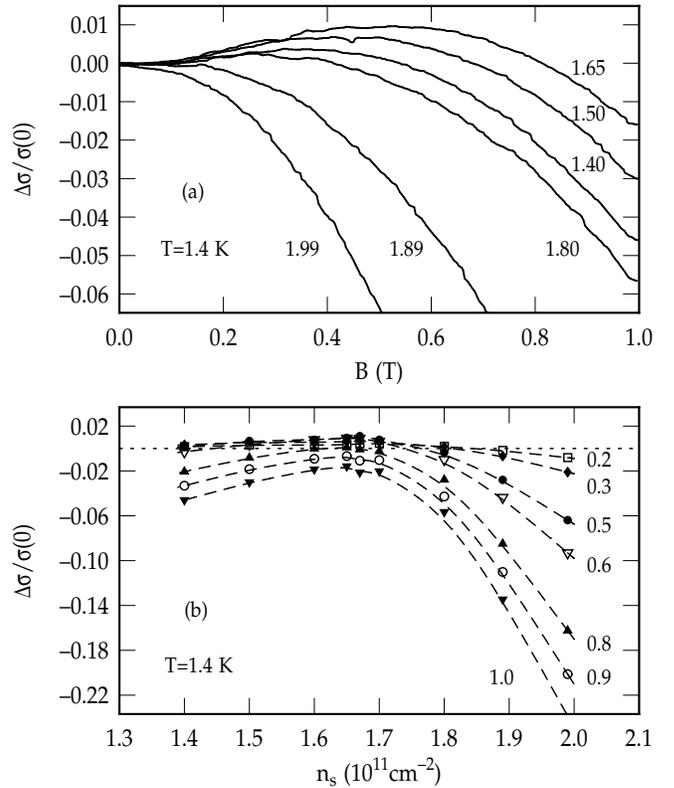}\vspace{5pt}
\caption{Magnetoconductance for sample \#1 with $V_{sub}=-9$~V and
$n_{c}=1.65\times 10^{11}$cm$^{-2}$.  (a) The carrier densities corresponding
to different curves are shown on the plot in units of $10^{11}$cm$^{-2}$.  (b)
The magnetic fields corresponding to different symbols are shown on the plot in
units of Tesla.  Dashed lines are guides to the eye.
\label{dhva1mc}}
\end{figure}

We have performed careful magnetoconductance measurements in 
perpendicular magnetic fields $B$ of up to 1~T.  Fig.~\ref{dhva1mc}(a) shows
magnetoconductance (MC), defined as $\Delta\sigma/\sigma (0)=[\sigma (B)-
\sigma (0)]/\sigma (0)$, for sample \#1 with $V_{sub}=-9$~V, for several 
carrier densities in the critical region.  As the (zero-field) critical
density $n_c$ is approached from the metallic side, the
positive MC develops gradually, reaches a maximum at $B\sim 0.5$~T when 
$n_{s}=n_c$, and then it disappears gradually on the insulating side.  
Exactly the same behavior is found in high-mobility mesoscopic 
Si MOSFETs~\cite{Li} at much lower temperatures, down to 40~mK.  At higher 
$B$, and for $n_s$ outside this narrow critical region, MC is always negative.
The dependence of MC on $n_s$ is displayed clearly in Fig.~\ref{dhva1mc}(b) for
several magnetic fields.  MC exhibits a pronounced maximum at $n_{s}=n_{c}$.

Obviously, there are positive and negative contributions to the measured MC.
It is well established that the positive MC can result from quantum 
interference effects in both weakly~\cite{LR} and strongly~\cite{Shkl,Sivan} 
disordered systems.  The spin-dependent part of the electron-electron 
interaction~\cite{LR} gives rise to negative MC, at least in the weakly
disordered regime.  This has been well established in doped 
semiconductors~\cite{snezanal}, and large negative MC has been seen in 
high-mobility MOSFETs for $B$ parallel to the current plane~\cite{dima}.
In all cases, one expects $|\Delta\sigma /\sigma (0)|\propto B^{2}$
in the low-field limits, 
becoming weaker for $B$ higher than some characteristic field.  In case of 
weak localization, the characteristic field is reached when the Landau orbit 
size becomes comparable to the thermal length, i.~e. when 
$2eB_0/\hbar\sim kT/\hbar D$, where $D$ is the diffusion constant~\cite{LR}, so
that $B_0\sim 0.3$~T for our samples.
In our experiment, MC changes sign and becomes
negative at sufficiently high $B$ for all $n_s$ in the critical region
($|\delta_{n}|\leq 0.1$).  We, therefore, conclude that the positive 
contribution is no longer in the low-field limit, consistent with the estimate
of $B_0\sim 0.3$~T.
On the other hand, there is a negative, predominantly quadratic 
$B$-dependence at the highest fields, indicating that the negative
contribution to the MC is still in the low-field regime.  This is consistent
with the estimated characteristic field for the MC due to spin splitting in an
interacting electron gas, given by the condition that the Zeeman energy is
comparable to the thermal energy, i.~e. $g\mu_{B}B\sim kT$~\cite{LR}, and for
$g\sim 2$~\cite{gabriele}, $g\mu_{B}B\sim 1.4$~K at $B=1$~T.

\begin{figure}[t]
\epsfxsize=3.4in \epsfbox{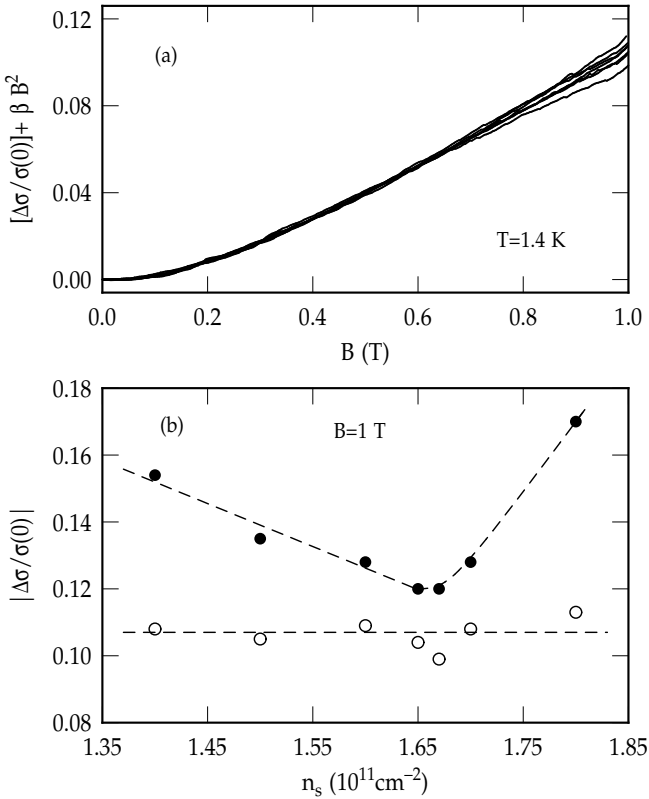}\vspace{5pt}
\caption{(a) The positive contribution [$\alpha f(B)$] to magnetoconductance 
for all $1.40\leq n_s (10^{11}$cm$^{-2})\leq 1.80$.
(b) Magnetoconductance vs. $n_s$ for $B=1$~T.  Open symbols represent the
positive contribution, and solid symbols represent the negative one.
Dashed lines are guides to the eye.
\label{dhva1mc2}}
\end{figure}

In order to understand the puzzling $n_s$-dependence of MC 
[Fig.~\ref{dhva1mc}(b)], it is desirable to disentangle the two contributions.
Obviously, the MC in our samples can be described by
\begin{equation}
\label{eq3}
\Delta\sigma /\sigma (0)=\alpha f(B) - \beta B^{2}, 
\end{equation}
with $\alpha , \beta >0$.  We find that, for $B>0.3$~T, $[\Delta\sigma 
/\sigma (0)]/B^2$ has the same form, 
up to a constant ($\beta$), for all $n_s$.  Fig.~\ref{dhva1mc2}(a)
shows that different $\alpha f(B)$ curves are indeed indistinguishable to 
within the scatter of data and, therefore, independent of $n_s$.  In this way,
we have been able to determine the change in $\beta$ as $n_s$ is varied,
independent of the choice of $f(B)$.  In order to obtain the absolute values of
$\beta$ to show in Fig.~\ref{dhva1mc2}, we have used 
$f(B)=B^{3/2}$~\cite{comment}.
Fig.~\ref{dhva1mc2}(b) presents both the positive ($\alpha f(B)$: open symbols)
and the negative contribution ($\beta B^2$: solid symbols) as a function of
$n_s$ for $B=1$~T.
The positive MC due to quantum interference does not depend on $n_s$ within 
the scatter of data.  On the other hand, the negative contribution,
i.~e. $\beta (n_{s})$, has a strong minimum at the metal-insulator 
transition.  

In summary, we have shown that the metallic phase can be created in a 2DEG
by increasing its mobility, consistent with the assertion that the
2D MIT is driven by electron-electron interactions.
We have also shown that the negative part of the MC, determined by the 
spin-dependent part of the interaction, decreases sharply at the transition.
We point out striking similarities between transport properties of a 2DEG in Si
MOSFETs and those observed in Si:B, which is a 3D system: (a) the
$T$-dependence is qualitatively the same, i.~e. $\sigma$ increases with 
decreasing $T$ in the metallic phase~\cite{myriam}; (b)  magnetoconductance,
which is negative in Si:B, depends strongly on the carrier concentration and 
shows a dramatic decrease at the transition~\cite{snezana}.  We recall that 
such anomalous behavior has been attributed to electron-electron interactions 
in the case of Si:B.  Our experiment suggests the possibility that 
electron-electron interactions may play a similar role in both 2D and 3D 
systems near the metal-insulator transition.

The authors are grateful to V. Dobrosavljevi\'{c} for useful discussions.  
This work was supported by NSF Grant No. DMR-9510355.


\vspace{-12pt}

\begin{references}

\bibitem{gang} E.\ Abrahams, P.W.\ Anderson, D.C.\ Licciardello, and
T.V.\ Ramakrishnan, Phys.\ Rev.\ Lett.~{\bf 42}, 673 (1979).

\bibitem{weakloc} D.J.\ Bishop, D.C.\ Tsui, and R.C.\ Dynes, Phys. Rev.
Lett.~{\bf 44}, 1153 (1980); M.J.\ Uren, R.A.\ Davies, and M.\ Pepper, 
J.\ Phys.\ C: Solid St. Phys.~{\bf 13}, L985 (1980).

\bibitem{Krav} S.V.\ Kravchenko, G.V.\ Kravchenko, J.E.\ Furneaux, V.M.\
Pudalov, and M.\ D'Iorio, Phys. Rev. B~{\bf 50} 8039 (1994);
S.V.\ Kravchenko, Whitney E.\ Mason, G.E.\ Bowker, J.E.\ 
Furneaux, V.M.\ Pudalov, and M.\ D'Iorio, Phys.\ Rev.\ B~{\bf 51}, 7038 (1995);
S.V.\ Kravchenko, D.\ Simonian, M.P.\ Sarachik, Whitney Mason, and J.E.\
Furneaux, Phys.\ Rev.\ Lett.~{\bf 77}, 4938 (1996).

\bibitem{dp} Dragana Popovi\'{c} and S.\ Washburn, preprint cond-mat/9612021.

\bibitem{newgang} V.\ Dobrosavljevi\'{c}, E.\ Abrahams, E.\ Miranda, and Sudip
Chakravarty, preprint cond-mat/9704091.

\bibitem{AFS} See T.\ Ando, A.B.\ Fowler, and F.\ Stern, Rev.\ Mod.\ 
Phys.~{\bf 54}, 437 (1982).

\bibitem{Alan} A.B.\ Fowler, Phys.\ Rev.\ Lett.~{\bf 34}, 15 (1975).

\bibitem{Li} Kuo-Ping Li, Dragana Popovi\'{c}, and S.\ Washburn (unpublished).

\bibitem{LR} P.A.\ Lee and T.V.\ Ramakrishnan, Rev.\ Mod.\ Phys.\ {\bf 57}, 287
(1985).

\bibitem{Shkl} V.L.\ Nguyen, B.Z.\ Spivak, and B.I.\ Shklovskii, Sov.\ Phys.\
JETP {\bf 62}, 1021 (1985).

\bibitem{Sivan} U.\ Sivan, O.\ Entin-Wohlman, and Y.\ Imry, Phys.\ Rev.\ 
Lett.\ {\bf
60}, 1566 (1988); O.\ Entin-Wohlman, Y.\ Imry, and U.\ Sivan, 
Phys.\ Rev.\ B {\bf 
40}, 8342 (1989).

\bibitem{snezanal} S.\ Bogdanovich, P.\ Dai, M.P.\ Sarachik, and V.\ 
Dobrosavljevi\'{c}, Phys.\ Rev.\ Lett.\ {\bf 74}, 2543 (1995).

\bibitem{dima} D.\ Simonian, S.V.\ Kravchenko, and M.P.\ Sarachik, preprint
cond-mat/9704071.

\bibitem{gabriele} g-factor may be enhanced up to $g\sim 4$ at low $n_s$: see
Sudhakar Yarlagadda and Gabriele F.\ Giuliani, Phys.\ Rev.\ B {\bf 49}, 14188
(1994).

\bibitem{comment} The form of the function cannot be determined precisely from
the existing data, but our qualitative conclusions do not depend on the choice
of $f(B)$.

\bibitem{myriam} P.\ Dai, Y.\ Zhang, and M.P.\ Sarachik, Phys.\ Rev.\ Lett.\ 
{\bf 66}, 1914 (1991).

\bibitem{snezana} S.\ Bogdanovich, P.\ Dai, M.P.\ Sarachik, 
V.\ Dobrosavljevi\'{c}, and G.\ Kotliar, Phys.\ Rev.\ B {\bf 55}, 4215 (1997).

\end{references}
\end{document}